\documentclass[aps,prl,twocolumn,floats]{revtex4}

\usepackage{amssymb}
\usepackage{amsmath}
\usepackage{graphicx}
\usepackage{bm}
\usepackage{mdframed}
\usepackage{color}

\begin{document}

\bibliographystyle{prsty}
\author{X. Z. Chen,$^{1}$ R. Zarzuela,$^{2}$ J. Zhang,$^{3}$ C. Song,$^{1,*}$ X. F. Zhou,$^{1}$ G. Y. Shi,$^{1}$ F. Li,$^{1}$ H. A. Zhou,$^{4}$ W. J. Jiang,$^{4}$ F. Pan,$^{1}$ and Y. Tserkovnyak$^{2}$}
\affiliation{$^{1}$Key Laboratory of Advanced Materials (MOE), School of Materials Science and Engineering, Tsinghua University, Beijing 100084, China \\
$^{2}$Department of Physics and Astronomy, University of California, Los Angeles, California 90095, USA \\
$^{3}$School of Physics and Wuhan National High Magnetic Field Center, Huazhong University of Science and Technology, Wuhan 430074, China \\
$^{4}$State Key Laboratory of Low-Dimensional Quantum Physics and Department of Physics, Tsinghua University, Beijing 100084, China}

\email{songcheng@tsinghua.edu.cn}

\begin{abstract}
We investigate the current-induced switching of the N\'{e}el order in NiO(001)/Pt heterostructures, which is manifested electrically via the spin Hall magnetoresistance. Significant reversible changes in the longitudinal and transverse resistances are found at room temperature for a current threshold lying in the range of $10^{7}$ A/cm$^{2}$. The order-parameter switching is ascribed to the antiferromagnetic dynamics triggered by the (current-induced) antidamping torque, which orients the N\'{e}el order towards the direction of the writing current. This is in stark contrast to the case of antiferromagnets such as Mn$_{2}$Au and CuMnAs, where fieldlike torques induced by the Edelstein effect drive the N\'{e}el switching, therefore resulting in an orthogonal alignment between the N\'{e}el order and the writing current. Our findings can be readily generalized to other biaxial antiferromagnets, providing broad opportunities for all-electrical writing and readout in antiferromagnetic spintronics.
\end{abstract}

\title{Antidamping torque-induced switching in biaxial antiferromagnetic insulators}

\maketitle

{\it Introduction}.|Antiferromagnets exhibit ultrafast spin dynamics with characteristic frequencies in the THz range, produce negligible stray fields and are robust against magnetic perturbations, offering prospects for the design of reliable high-density memories with  fast operation speeds \cite{Jungwirth-NN2016,Wadley-Science2016,Olejnik-NatComm2017,Chen-NatComm2017}. Thus, it is crucial to develop new avenues for manipulating the antiferromagnetic (AFM) order, of particular interest being electrical methods based on the spin-transfer effect. Spin-orbit torques (SOT) can be split into two generic classes, namely fieldlike and antidamping torques: $\partial_{t}\bm{m}|_{\textrm{FL}}\sim\bm{m}\times\bm{p}$ \cite{Wadley-Science2016,Olejnik-NatComm2017} and $\partial_{t}\bm{m}|_{\textrm{AD}}\sim\bm{m}\times(\bm{m}\times\bm{p})$ \cite{Liu-Science2012,Sinova-RMP2015}, respectively, where $\bm{m}$ denotes the orientation of a magnetic sublattice and $\bm{p}$ represents the nonequilibrium spin polarization. For bipartite antiferromagnets, if $\bm{p}$ is the same on both sublattices, the fieldlike torque is equivalent to that exerted by a uniform magnetic field, forcing the N\'{e}el order to orient perpendicular $\propto\bm{p}$. While this could mimic the antiferromagnetic spin-flop transition, the corresponding critical field is in the range of tens of Tesla for conventional antiferromagnets, which would translate into prohibitively large charge currents.

Recently, Wadley et al. \cite{Wadley-Science2016} reported a current-induced switching of the AFM order in CuMnAs, setting a milestone for the manipulation of the staggered order parameter. In such material, where the breaking of inversion symmetry occurs at the sublattice level, opposite spin polarizations are induced in the two (inversion-partner) sublattices, $\bm{p}_{1}=-\bm{p}_{2}$, via the Edelstein effect. The ensuing fieldlike torque can reorient the N\'{e}el order towards the (current-induced) effective field $\propto\bm{p}_{1}-\bm{p}_{2}$, whenever this field overcomes the threshold given by the in-plane anisotropy (several tens of Oersted), as dictated by the effective energetics of the order parameter. Even though this SOT seems attractive for the low-current switching of the N\'{e}el order, the underlying inverse spin-galvanic mechanism imposes stringent requirements on the crystallographic structure and quality of the antiferromagnet, namely global centrosymmetry plus broken sublattice inversion symmetry. At present, only a few antiferromagnets such as CuMnAs and Mn$_{2}$Au can meet this demand \cite{Wadley-Science2016,Olejnik-NatComm2017,Zelezny-PRL2014}.

Antidamping torques have also been proposed to trigger oscillations and even the switching of AFM moments regardless of the crystal symmetries \cite{Zelezny-PRL2014,Gomonay-PRB2010,Cheng-PRL2016,Hals-PRL2011,Zarzuela-PRBRC2017}. This antidamping switching mechanism could be extended to a wide range of antiferromagnets with biaxial anisotropy, allowing for the extensive materials usage. However, no conclusive experimental observation of the antidamping torque-induced switching has been reported yet in AFM/HM (heavy metal) heterostructures. Moriyama et al. \cite{Moriyama-2017} reported very recently the control by SOT of AFM moments in Pt/NiO(111)/Pt trilayers. While the antidamping torques were invoked as the driving force behind this magnetization switching, the symmetries of the structure instead seem to point to the fieldlike scenario. Here, we investigate the switching of AFM moments mediated by antidamping torques in the biaxial NiO(001)/Pt heterostructure. We utilize the spin Hall magnetoresistance (SMR) as a probe for the (dynamics of the) magnetic moments of this AFM insulator \cite{Nakayama-PRL2013,SMR,Yokoyama-PRB2014,Manchon-PSS2017,Han-PRB2014}, which is well known to display a negative SMR signal \cite{Hoogeboom-APL2017,Hou-PRL2017,Lin-PRL2017}. We observe significant changes in the longitudinal and transverse resistances when current pulses are applied along the orthogonal easy axes of NiO, corresponding to the N\'{e}el switching towards the direction of the current.

\begin{figure}[t]
\begin{center}
\includegraphics[width=1.0\linewidth]{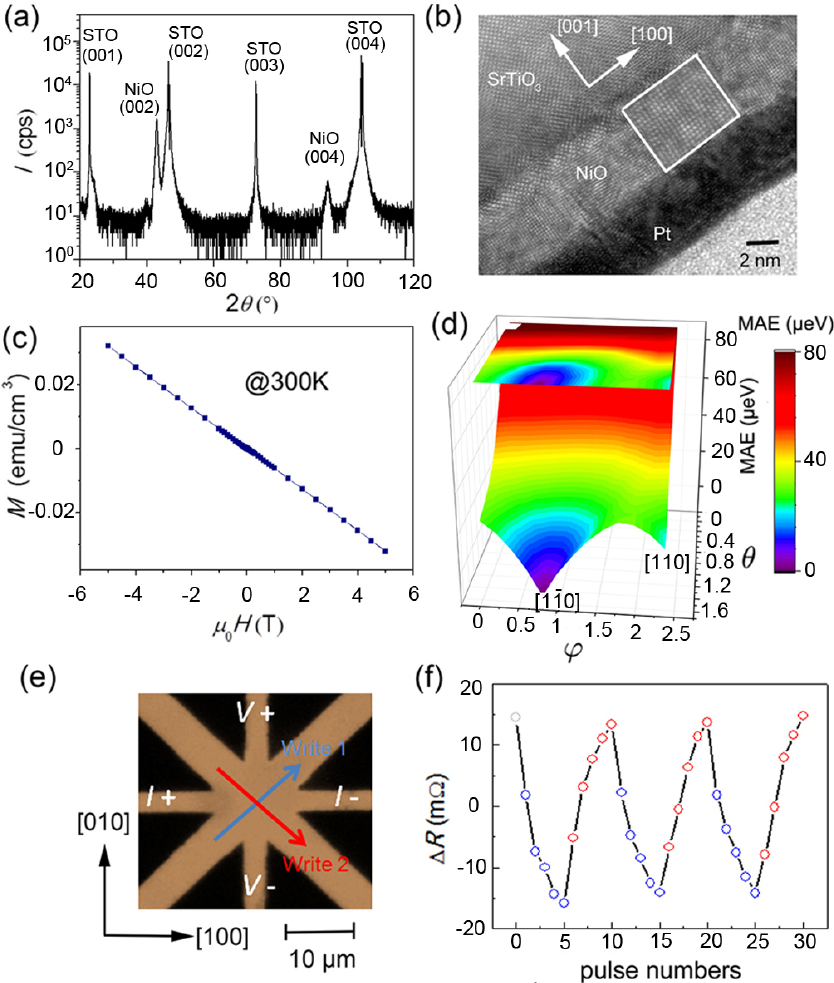}
\caption{(a) X-ray diffraction pattern, (b) high resolution TEM images and (c) room-temperature hysteresis loop of the SrTiO$_{3}$(001)/NiO/Pt(5 nm) heterostructure. The inset of subfigure (b) shows the image of the NiO lattice, which has been magnified twice. (d) Spatial (angular) dependence of the magnetocrystalline anisotropy energy for the tetragonal NiO. The easy axis points towards $[1\bar{1}0]$, while the local easy axis is $[110]$. (e) Optical image of a NiO/Pt 8-terminal device and schematic of current-driven switching measurements. (f) Application of  writing currents along $[110]$ and $[1\bar{1}0]$ yields reversible changes in the Hall resistance that can be detected at $T=300$ K. Hall resistances are measured  $10$ s after applying the current pulses.}
\label{Fig1}
\end{center}
\vspace{-0.5cm}
\end{figure}

{\it Experimental results.}|The in-plane biaxial NiO films were obtained by deposition under compressive strain on SrTiO$_{3}$ (STO) substrates ($a_{\textrm{NiO}}\sim0.4177$ nm and $a_{\textrm{STO}}\sim0.3905$ nm) by using magnetron sputtering at 473 K, and capped with 5 nm-thick Pt after cooling down to room temperature. Afterwards, the NiO(5 nm)/Pt(5 nm) bilayers are patterned to the desired geometry by using electron beam lithography and Ar ion etching.

Figure \ref{Fig1}(a) shows the X-ray diffraction pattern of the NiO(100 nm)/Pt(5 nm) stack on STO(001), where only (002) and (004) NiO peaks and the substrate peaks are visible in the range from 20$^{\circ}$ to 120$^{\circ}$, therefore reflecting the quasi-epitaxial growth of the NiO. The parameter $c\sim0.4205$ nm, deduced from the (004) peak at 94.2$^{\circ}$, confirms that NiO on STO is elongated out-of-plane or, equivalently, constrained within the plane. As shown in Fig. \ref{Fig1}(b), the cross-sectional transmission electron microscopy (TEM) images of the STO/NiO(5 nm)/Pt(5 nm) heterostructure illustrate the high sample quality of NiO, with the stack relationship of NiO(001)[100]$\parallel$STO(001)[100] and a sharp interface between the NiO and Pt. The NiO lattice is highlighted in the inset panel, magnified twice as compared to the whole TEM image. Magnetization measurements were carried out at $300$ K in a Superconducting Quantum Interference Device to reveal the AFM order of NiO. Magnetic fields were swept from $-5$ T to $5$ T, and only the diamagnetic background of the STO substrate was observed [Fig. \ref{Fig1}(c)].

Based on the lattice parameters deduced from XRD, first-principles calculations of the magnetocrystalline anisotropy (MCA) of the tetragonal NiO were performed. Figure \ref{Fig1}(d) depicts the spatial dependence of the MCA energy in terms of the angles $\theta,\varphi$ (spherical representation of the order parameter). Note that $[1\bar{1}0]$ and $[001]$ are the easy and hard axes of the NiO sample, respectively, while a second (local) easy axis aligns along $[110]$. The easy plane rotates from $(111)$ to $(001)$ due to the fourfold in-plane constraint imposed by the substrate. Also visible is that the NiO MCA is greatly enhanced as compared to the bulk \cite{SM}. An optical image of a NiO/Pt 8-terminal device and the schematic of the current-driven switching measurement are depicted in Fig. \ref{Fig1}(e). As the SMR measurements in Fig. \ref{Fig1}(f) show, two orthogonal orientations (easy axes) of the N\'{e}el order can be reversibly written by applying current pulses at room temperature. For this experiment, the writing current channels along the two orthogonal directions $[110]$ and $[1\bar{1}0]$ of the NiO film, Write 1 and 2, respectively, allow for the control of the spin polarization of the injected spin current. The $[100]$ and $[010]$ channels are used for probing AFM moments via the longitudinal and transverse SMR, where current and voltage ($I$ and $V$) stand for the detection. The change in the Hall resistance is $\sim30$ m$\Omega$, corresponding to a ratio of $\sim$0.05\% when divided by the longitudinal resistance. An inspection of the figure shows the gradual change in the Hall resistance with consecutive current pulses, indicating the multidomain switching behavior.

\begin{figure}[t]
\begin{center}
\includegraphics[width=1.0\linewidth]{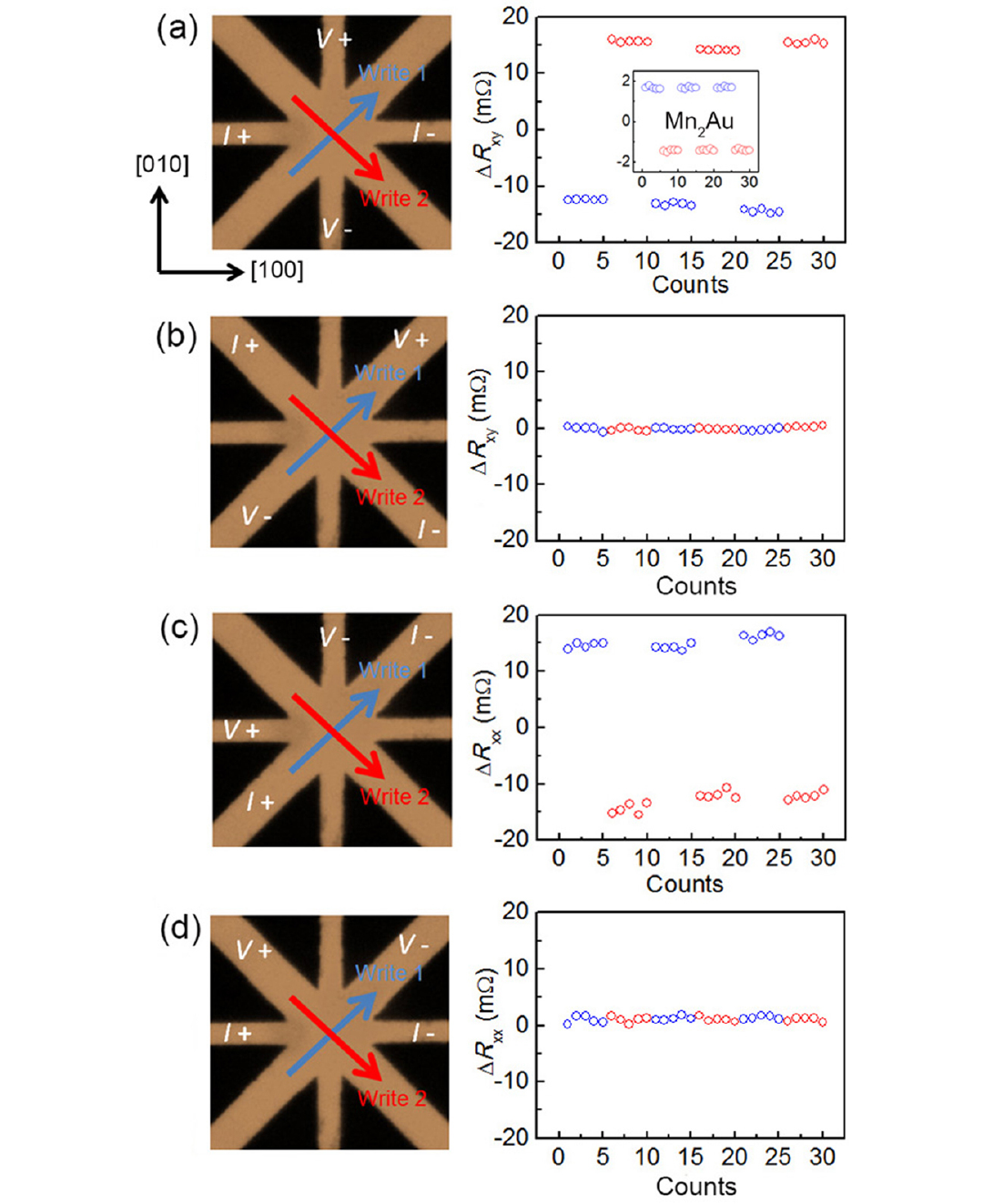}
\caption{Detection of the N\'{e}el order switching for transverse [(a)-(b)] and longitudinal [(c)-(d)] geometries. Blue (red) arrows indicate the current being applied along the writing channel $[110]$ $([1\bar{1}0])$. $I$ and $V$ represent the readout scheme. The writing current density used is $4\times10^{7}$ A/cm$^{2}$. Each circle in the right column results from the application of five 1-ms pulses with delay time of 5 s, where the color corresponds to the writing channel considered.}
\label{Fig2}
\end{center}
\vspace{-0.5cm}
\end{figure}

Figure \ref{Fig2} illustrates the current-induced N\'{e}el switching in the tetragonal (biaxial) NiO. In Figs. \ref{Fig2}(a)-(d) both the magnitude and direction of the writing current are identical, but the configurations of the detected $I$ and $V$ are different, as highlighted in the left column. Five 1-ms current pulses with the delay time of 10 s were applied along the writing channels. The current amplitude used was $4\times10^{7}$ A/cm$^{2}$, comparable to  the ferrimagnetic system \cite{Avci-NatMat2017,Peng-NatCommun2016}, and a small probe dc current of $4\times10^{5}$ A/cm$^{2}$ was applied along $[100]$. Figures \ref{Fig2}(a) and (b) show the detection of the AFM moments in the transverse geometry. Two distinct values for the transverse resistance were observed reversibly, see Fig. \ref{Fig2}(a), the total change in resistance being about 30 m$\Omega$. Since $\rho_{xy}^{\textrm{sH}}=\Delta\rho\sin2\theta$ \cite{SMR}, the change in the Hall resistance is maximal (in absolute value) at the relative angles of $\theta=45,135^{\circ}$ between the probe current and the N\'{e}el order; therefore, writing protocols along $[110]$ (high resistance) and $[1\bar{1}0]$ (low resistance) correspond to $\theta=45^{\circ}$ and $\theta=135^{\circ}$, respectively, i.e. the AFM order switches towards the direction of the writing current. In contrast, the inset panel shows the current-induced N\'{e}el switching in our Mn$_{2}$Au control samples for the identical experimental configuration, where the anisotropic magnetoresistance of Mn$_{2}$Au and the SMR of NiO/Pt possesses the same in-plane symmetry \cite{Nakayama-PRL2013,SMR}. The concomitant sign of the changes in the Hall resistance is opposite to that of the NiO/Pt scenario, because the (fieldlike) Edelstein SOT would switch the N\'{e}el order towards the spin polarization direction, which is transverse to the writing current, similar to the CuMnAs case \cite{Wadley-Science2016,Zelezny-PRL2014}.

For the sake of completeness we considered a different Hall detection scheme, as shown in Fig. \ref{Fig2}(b). Insignificant changes in the transverse resistance were measured after running the same current pulse protocol, which correspond to the angles $\theta=0^{\circ}$ (red circles) and $\theta=90^{\circ}$ (blue circles). This is in accordance with the aforementioned expression for $\rho_{xy}^{\textrm{sH}}$. Furthermore, we studied the symmetry of SMR for the longitudinal geometry, see Figs. \ref{Fig2}(c) and (d). The longitudinal resistance increases when the writing current is parallel to the probe current. Theory of SMR predicts $\rho_{xx}^{\textrm{sH}}=\rho_{0}+\Delta\rho\cos2\theta$ \cite{SMR}, so that larger values of the longitudinal resistance indicate the parallel alignment of the AFM moments with the probe current. Again, the control experiment depicted in Fig. \ref{Fig2}(d) shows negligible changes in the longitudinal resistance, which agrees with the theoretical predictions. Therefore, we conclude that the N\'{e}el order switches towards the direction of the writing current. Note that both transverse and longitudinal resistances were measured one minute after the current pulses to minimize the Joule heating.

{\it Model.}|Dynamics of the N\'{e}el order, defined by $\bm{l}=(\bm{m}_{1}-\bm{m}_{2})/2$, with $\bm{m}_{1},\bm{m}_{2}$ being the sublattice spin fields, are described by the equation of motion
\begin{equation}
\label{eq1}
s\chi\partial_{t}^{2}\bm{l}\times\bm{l}=\bm{H}_{\textrm{an}}\times\bm{l}+\alpha\bm{l}\times\partial_{t}\bm{l}+(\vartheta J/s)\bm{l}\times(\bm{l}\times\bm{p}),
\end{equation}
along with the normalization condition $\bm{l}^{2}\equiv1$. Here, $s$ is the saturated spin density, $\chi$ is the (transverse) spin susceptibility, $\bm{H}_{\textrm{an}}=-\delta_{\bm{l}}\mathcal{E}_{\textrm{MCA}}/s$ is the anisotropy field, $\alpha=0.007$ is the Gilbert damping constant due to spin pumping in our 5-nm film \cite{Tserkovnyak-PRL2002,Khymyn-SRep2017}, which is also the value used in Ref. \onlinecite{Cheng-PRL2016}, and $\vartheta$ is a phenomenological parameter characterizing the charge-to-spin current conversion at the NiO/Pt interface, which can be approximated by $\vartheta\sim0.1\hbar/2e$. The last term of the right-hand side represents the current-induced antidamping torque, which vanishes for $\bm{l}\parallel\bm{p}$. Contrary to the ferromagnetic case, this configuration of the N\'{e}el order is unstable against perturbations, as predicted by the stability analysis of Eq. \eqref{eq1} \cite{Zarzuela-PRBRC2017,Hirch-Book2004}: Thermal fluctuations will induce slight deviations of the N\'{e}el order from the ground state $\bm{l}\parallel\bm{p}$, which, above some current threshold, will evolve in time towards the plane perpendicular to the spin polarization, $\bm{l}\perp\bm{p}$. Note, however, that for the case of uniaxial anisotropy (NiO on MgO substrate) the N\'{e}el order will switch back to the (only) easy axis when the applied current is switched off, irrespective of its intensity \cite{SM}. This is not the case for the (in-plane) biaxial anisotropy, for which writing currents along orthogonal (easy-axis) directions allow for the reversible manipulation of the AFM moments between the easy axes. It is worth remarking that reactive fieldlike torques $\propto\bm{l}\times\bm{p}$ are forbidden if the sublattice symmetry $\{\bm{l}\rightarrow-\bm{l}\}$ is preserved.

\begin{figure}[t]
\begin{center}
\includegraphics[width=1.0\linewidth]{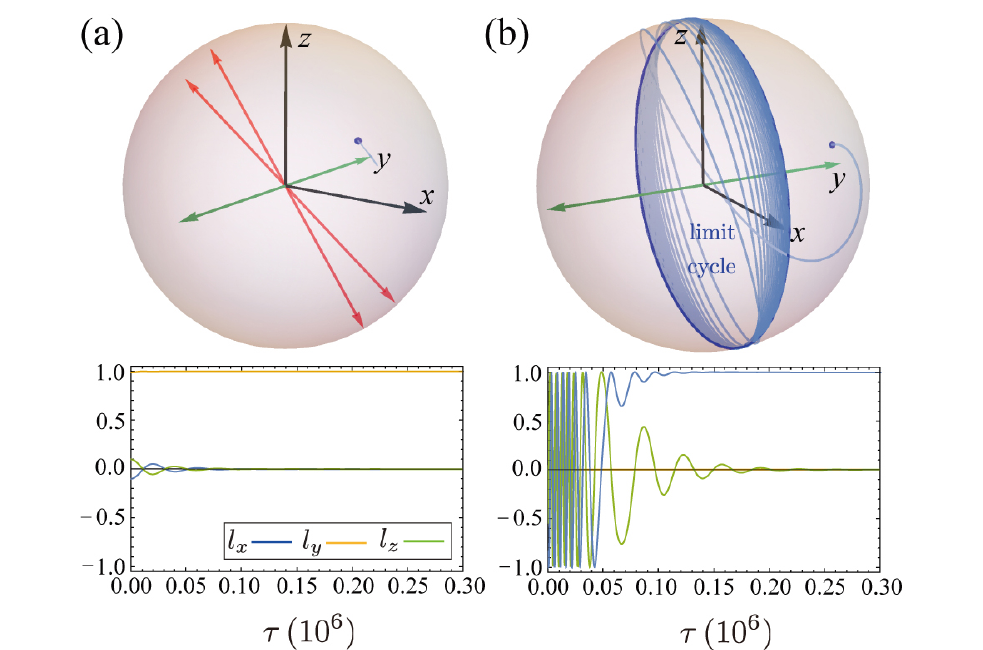}
\caption{Monodomain simulation of the N\'{e}el switching mediated by antidamping torques for a charge current injected along the $x$ axis. Panels (a) and (b) show the fixed points and limit cycles of the dynamical system in the N\'{e}el-order sphere (up) and the time evolution of the N\'{e}el order (bottom) for $J_{x}=4.9\times10^{8}$ A/cm$^{2}$ and $J_{x}=2.2\times10^{9}$ A/cm$^{2}$, respectively. Blue dots represent the initial configuration $\bm{l}=(-0.1, 0.99, 0.1)$ for the N\'{e}el order and blue trajectories describe the complete time evolution of the order parameter towards (a) the $y$ axis and (b) the limit cycle. Red arrows denote the fixed points of the dynamical system, along with the easy axis (green arrows). The limit cycle in (b) lies within the $xz$ plane. After turning off the current, the N\'{e}el order switches towards the $x$ axis (local minimum of energy) above the critical threshold $J_{x,c}=5.8\times10^{8}$ A/cm$^{2}$.}
\label{Fig3}
\end{center}
\end{figure}

Further insight into this switching process can be gained through monodomain simulations of the N\'{e}el dynamics driven by antidamping torques. Zero-temperature values of the anisotropy constants are obtained by fitting the MCA energy calculated by the DFT+U method \cite{SM} to the expression $\mathcal{E}_{\textrm{MCA}}[\bm{l}]=K_{0}+(K_{2\perp}/2)l_{z}^{2}+(K_{4\parallel}/4)l_{x}^{2}l_{y}^{2}+K_{2\parallel}(l_{x}^{2}-l_{y}^{2})$, which contains all (dominant) terms allowed by the tetragonal symmetry up to 4-th order in the N\'{e}el field. Current pulses of amplitude $4\times10^{7}$ A/cm$^{2}$ and 1-ms width would heat the sample to $T=350$ K \cite{SM}. Thermal effects can be incorporated into our model by invoking Bloch's $T^{2}$ law for the sublattice spin density $\mathfrak{s}$ in antiferromagnets \cite{Kubo-PR52,Oguchi-PR60} and the power law $K_{l\parallel},K_{l\perp}\propto\mathfrak{s}^{l(l+1)/2}$ for the MCA coefficients \cite{Callen-1966}. The resultant estimates for the anisotropy constants are $K_{2\perp}=3.8\times10^{-8}$ J/cm$^{2}$, $K_{2\parallel}=4.3\times10^{-9}$ J/cm$^{2}$ and $K_{4\parallel}=1.4\times10^{-9}$ J/cm$^{2}$. The ground N\'{e}el state thus lies along the $y$ direction (absolute minimum of the MCA energy) within the easy basal ($xy$) plane. Note that in the above expression the frame of reference has been rotated so that $\hat{y}\,(\hat{x})$ is the (local) easy axis. We take the writing current along the $x$ axis in what follows. Stability analysis of the fixed points $\bm{l}_{0}=\pm\hat{y}$ of Eq. \eqref{eq1} indicates that the N\'{e}el order along the easy axis is stable up to the critical current threshold
\begin{equation}
\label{eq2}
J_{x,c}^{\textrm{th}}=\frac{1}{2\vartheta}\sqrt{\mathcal{K}_{\perp}^{2}+\alpha^{2}\mathcal{K}_{\parallel}/\chi}\simeq5.8 \times10^{8}\textrm{ A/cm}^{2},
\end{equation}
where $\mathcal{K}_{\perp}=K_{2\perp}-K_{2\parallel}-K_{4\parallel}/2$ and $\mathcal{K}_{\parallel}=6K_{2\parallel}+2K_{2\perp}+K_{4\parallel}$ are effective anisotropy constants. Numerical integration of Eq. \eqref{eq1} leads to the N\'{e}el dynamics depicted in Fig. \ref{Fig3}: Transverse deviations from the easy-axis ground state die out below $J_{x,c}^{\textrm{th}}$, and therefore the N\'{e}el order eventually evolves back to the $y$ axis, as depicted in Fig. \ref{Fig3}(a). Above the current threshold, however, transverse excitations of the N\'{e}el order blow up until the staggered order parameter becomes confined to the $xz$ plane, see Fig. \ref{Fig3}(b), where its (limiting) dynamics take the form of a limit cycle. When the writing current is turned off (end of the pulse), no N\'{e}el dynamics occur below $J_{x,c}^{\textrm{th}}$, i.e., the order parameter remains aligned with the easy axis. Differently, above the critical threshold, the AFM order evolves in time to the minimum of MCA energy within the $xz$ plane, which corresponds to the local easy axis $\hat{x}$ in Fig. \ref{Fig3}(b).

\begin{figure}[t]
\begin{center}
\includegraphics[width=1.0\linewidth]{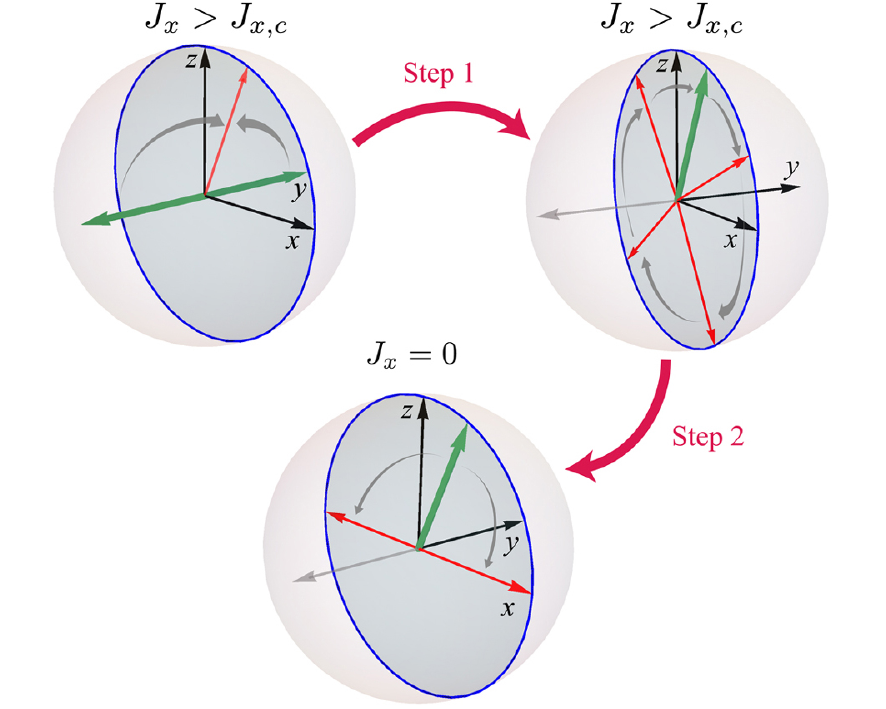}
\caption{Schematic of the two-step process involved in the antidamping torque-induced switching. Green (red) arrows depict initial (final) configurations for the N\'{e}el order. Blue circle depicts the limit cycle in the $xz$ plane.}
\label{Fig4}
\end{center}
\end{figure}

For writing currents along the $y$ axis, stability analysis of the fixed points $\pm\hat{x}$ of the dynamical system \eqref{eq1} yields their instability no matter what the injected current is. Irrespective of their steady state (either a stable fixed point or a limit cycle), the N\'{e}el order always evolves to the absolute minima of the MCA energy, $\pm\hat{y}$, when the current is turned off . Antidamping torque-induced switching of the order parameter in biaxial antiferromagnets is then summarized in Fig. \ref{Fig4}, which is described by a two-step process, namely (1) deviation from the easy axis and confinement to the plane perpendicular to the spin polarization, and (2) switching towards the local easy axis contained in the same plane.

{\it Discussion and conclusion}.|The critical threshold \eqref{eq2} is one order of magnitude larger than the measured threshold current. For ferromagnetic switching, similar overestimates of the threshold current within the monodomain treatment are well known \cite{Lee-PRB2K14,Zhang-APL2K15}. A switching process based on domain-wall nucleation and propagation can significantly lower the critical current. Indeed, the multistage character of switching in our pulse-sequence experiment suggests such multidomain mechanism for the order-parameter reorientation. While extending our theoretical picture towards this regime is left as an open problem, we remark that the monodomain treatment does appear to capture the physics qualitatively. More refined calculations and direct measurements of the magnetic anisotropies for our structures can further reconcile quantitative aspects. The reversible switching and detection reported here make biaxial AFM insulator/HM heterostructures promising candidates for all-electrical writing and readout of the N\'{e}el order.

{\it Acknowledgments}.|We thank Dr. R. Cheng, Dr. D. Z. Hou and Y. Z. Tan for helpful discussions. The authors in Tsinghua thank the support of Beijing Innovation Center for Future Chip (ICFC) and Young Chang Jiang Scholars Program. The work was partially supported by the National Key R\&D Program of China under Grant No. 2017YFB0405704, and the NSFC under Grant Nos. 51671110, 51571128 and 11704135. The work at UCLA was supported by NSF-funded MRSEC under Grant No. DMR-1420451 and NSF under Grant No. DMR-1742928.

{\it Note added}: During the preparation of this manuscript, we became aware of a report of the switching of AFM moments in Pt-sandwiched NiO films, with the critical current of $\sim 5\times10^{7}$ A/cm$^{2}$ \cite{Moriyama-2017}. Measured changes in the transverse Hall resistance were ascribed there to the time evolution of the N\'{e}el order towards $\bm{l}\parallel\bm{p}$, i.e., the order parameter being transverse to the writing current. This interpretation contrasts with the antidamping scenario, which could be explained that the trilayer Pt/NiO/Pt geometry might reduce the inversion asymmetry and thus diminish the relative importance of the antidamping torques. These issues need to be clarified in future studies on multilayers.

\end{document}